# Machine Learning Based Optical Thermometry Using Photoluminescence and Raman Spectra of Diamonds Containing SiV Centers

*Md Shakhawath Hossain,[†] Dylan G. Stone,[##] G. Landry,[##] Xiaoxue Xu,[††] Carlo Bradac[##,*] and Toan Trong Tran [†, *]*

[†] School of Electrical and Data Engineering, University of Technology Sydney, Ultimo, NSW, 2007, Australia.

[##] Department of Physics & Astronomy, Trent University, 1600 West Bank Dr., Peterborough, Ontario K9L 0G2, Canada

[††] School of Biomedical Engineering, University of Technology Sydney, Ultimo, NSW, 2007, Australia.

*Corresponding author: carlobradac@trentu.ca

*Corresponding author: trongtoan.tran@uts.edu.au

**Funding:** The Natural Sciences and Engineering Research Council of Canada (DGECR-2021-00234) and the Canada Foundation for Innovation (John R. Evans Leaders Fund #41173) are acknowledged. T. T. T acknowledges the financial support from the Australian Research Council (DE220100487, DP240103127).

## Abstract

Micro- and nanothermometry enable precise temperature monitoring and control at the micro/nanoscale, and have become essential diagnostic tools in applications ranging from high-power microelectronics to biosensing and nanomedicine. Most existing techniques rely on secondary micro-/nanothermometers that require the individual calibration of each sensor, ideally both off- and in-situ, before use. We present an alternative approach that overcomes this limitation by employing fluorescent diamonds containing silicon-vacancy centers, where the thermo-sensitive physical quantities are the centers' photoluminescence and the diamond host's Raman signals. The photoluminescence and Raman data are analyzed using two multi-feature regression algorithms that leverage a minimal number of calibration diamonds and temperature set points to predict the temperature of previously unseen diamonds. Using this approach, the models achieve accuracies as low as 0.7 K, resolutions down to 0.6 K·Hz$^{-1/2}$ and sensitivity as high as 0.04 K$^{-1}$. These correspond to improvements of roughly 70% (over threefold) in accuracy, 50% (twofold) in resolution and 567% (sevenfold) in sensitivity compared with traditional single-feature models. Our approach is particularly suited to applications where pre-deployment calibration of every thermosensor is impractical, and it is generalizable to any thermometry platform with two or more simultaneously measurable temperature-dependent observables.

These authors contribute equally to the project: *Md Shakhawath Hossain, Dylan G. Stone, G. Landry*.

**1. Introduction**

Fast and reliable micro/nanoscale optical thermometry remains in high demand due to the need for precise local temperature measurements, given that heat transfer at the micro/nanoscale can deviate significantly from classical Fourier's law.[1] Optical, probe-based contact thermometry has recently garnered significant attention in fields such as nanotechnology and materials science,[2-5] as well as biology and medicine.[6-8] This growing interest is mostly due to the marked ability of optical thermosensors to measure temperature at the micro/nanoscale, in a non-invasive manner and with minimal thermal load. In fluorescence optical thermometry, the temperature of target objects or the local environment is determined by measuring the temperature-dependent spectral properties of the probe—quantum dots,[9] diamond particles,[10-12] organic dyes,[13] upconversion nanoparticles,[14] etc. In particular, micro- or nanodiamonds hosting color centers have attracted considerable attention as thermal sensors. This stems from the fact that besides their excellent thermo-dependent photophysical properties,[15] they also display inherent biocompatibility[16] and mechanical robustness, making them ideal probes both for delicate (e.g. biological) and harsh (e.g. high-power electronic) environments. However, as secondary thermal sensors, these probes need calibration against reference temperatures,[17] a process that is both time-consuming and onerous, particularly when calibration of each individual thermal sensor is required. In addition, calibration should ideally be performed under both off-situ and in-situ conditions, since changes in the thermometers' surrounding environment can significantly and undesirably alter their optical behavior and generate thermal-equivalent noise (TEN).[18]

Here, we introduce two all-optical thermometry methods that exploit machine learning pattern recognition to address these limitations. The methods utilize fluorescent diamonds containing silicon-vacancy (SiV) centers as temperature sensors, and leverage multi-feature regression to predict the temperature of their surroundings by measuring thermally-driven changes in the sensors' spectroscopy. Although multi-feature regression analysis of fluorescent nanodiamonds has been explored previously,[19] the present approach introduces a few distinct innovations.

*i)* The proposed methods exploit the thermally-dependent, heterogeneous optical properties of diamond sensors by combining the photoluminescence (PL) of SiV centers and the Raman fluorescence of the diamond host. Access to multiple temperature-dependent features enhances the models' predictive power, while the strong temperature sensitivity of Raman fluorescence—widely regarded as the benchmark for high-sensitivity, high-resolution thermometry[20]—further boosts performance.

*ii)* Additionally, the proposed methods require minimal training data—only 3–4 diamonds and temperature calibration set points—to predict the temperature of any unseen diamond, with good generalized accuracy; in fact, ~26% to ~70% better than traditional single-feature (SF) models.

*iii)* Finally, the methods employ either an explicit (permutation-based) or implicit (regularization-based) feature selection strategy that automatically identifies the subset of features that minimize prediction errors. This makes the models highly adaptable to different user needs and instrumentation, enabling them to prioritize features based on the resolution, sensitivity, and accuracy available to them.

Our methods thus provide a flexible, data-efficient, and instrument-agnostic framework for nanoscale thermometry, specifically suited for practical deployment across diverse experimental platforms and in resource-constrained settings.

## 2. Results and Discussion
### 2.1. Experiment and Model

In our experiments, we used diamonds measuring (~1 μm) in diameter and containing a high concentration (~78 ppm)[21] of silicon-vacancy (SiV) centers. Briefly, the samples were prepared by drop-casting the diamonds onto a silicon dioxide substrate, followed by oxidative annealing in air at 550 °C to remove surface graphite-like carbon layers (cf. **§4.1 Methods**). Excitation and collection of photoluminescence (PL) and Raman scattering signals were carried out using a custom-built confocal microscope integrated with an open-loop, temperature-controlled cryostat (cf. **§4.2 Methods**).

Temperature estimation was performed by monitoring changes in the PL and Raman signals as the temperature of the sample was set up by the cryostat between 25 °C and 85 °C in 10 °C increments (see below). Spectra were recorded at each setpoint after thermal equilibrium had been reached (**Figure 1a–c**). At each temperature, up to nine physical quantities were simultaneously tracked. These are the intensity, peak position, and full width at half maximum (FWHM) of: *i)* the zero-phonon line (ZPL) and *ii)* phonon sideband (PSB) of the SiV centers, and *iii)* the Raman signal of the diamond host.

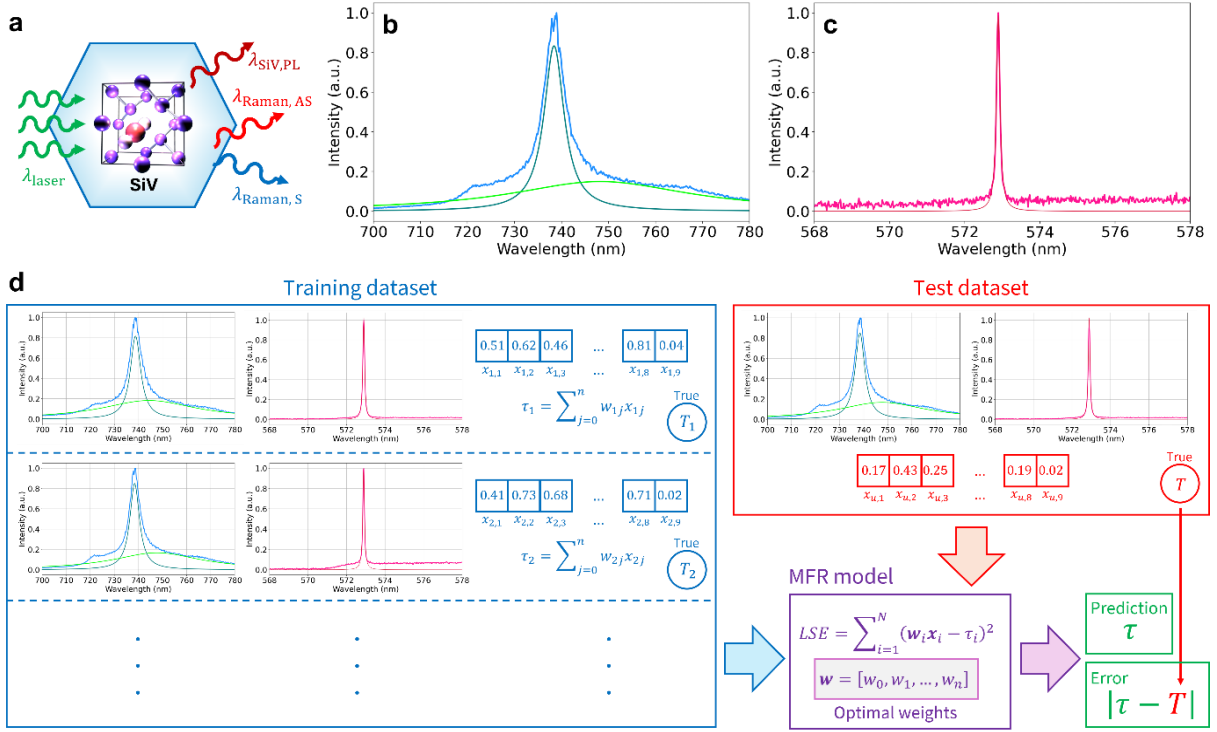

**Figure 1.** Data acquisition and multi-feature regression analysis. **a)** Schematic representation of the excitation of the diamonds and collection of the photoluminescence from the SiV centers and of the Raman signal from the diamond host. **b, c)** Photoluminescence (b) and Raman (c) spectra collected for a typical diamond at a reference temperature. Spectra were acquired with integration times of 1 s and 20 s for PL and Raman respectively, using a 532-nm continuous wave laser with 500 µW of power (measured at the back aperture of the objective). **d)** Flow chart showing the training and testing of the multi-feature regression (MFR) algorithm.

To determine the temperature from these measurements, we use two regression-based algorithms. The first one is a standard multi-feature regression (MFR) algorithm in which the target variable (i.e., the temperature) is modeled as a weighted linear combination of the identified temperature-dependent physical quantities—hereafter referred to as features (cf. **§4.3.1 Methods**). The second is a two-stage multi-feature regression (2SMFR) algorithm that uses partial least squares (PLS) regression, lasso regression and spline correction to estimate the temperature from the features and remove systematic biases (cf. **§4.3.2 Methods**). The ability to measure, simultaneously, several temperature-dependent physical quantities makes multi-feature regression-based algorithms a natural and attractive choice over traditional methods based on single-parameter tracking (e.g., ZPL wavelength or FWHM).[6-7, 17, 22] The underlying hypothesis is that multi-feature regression is less susceptible to parameter correlations and experimental fluctuations, as noise affecting individual features can be mitigated through averaging across multiple inputs.[23]

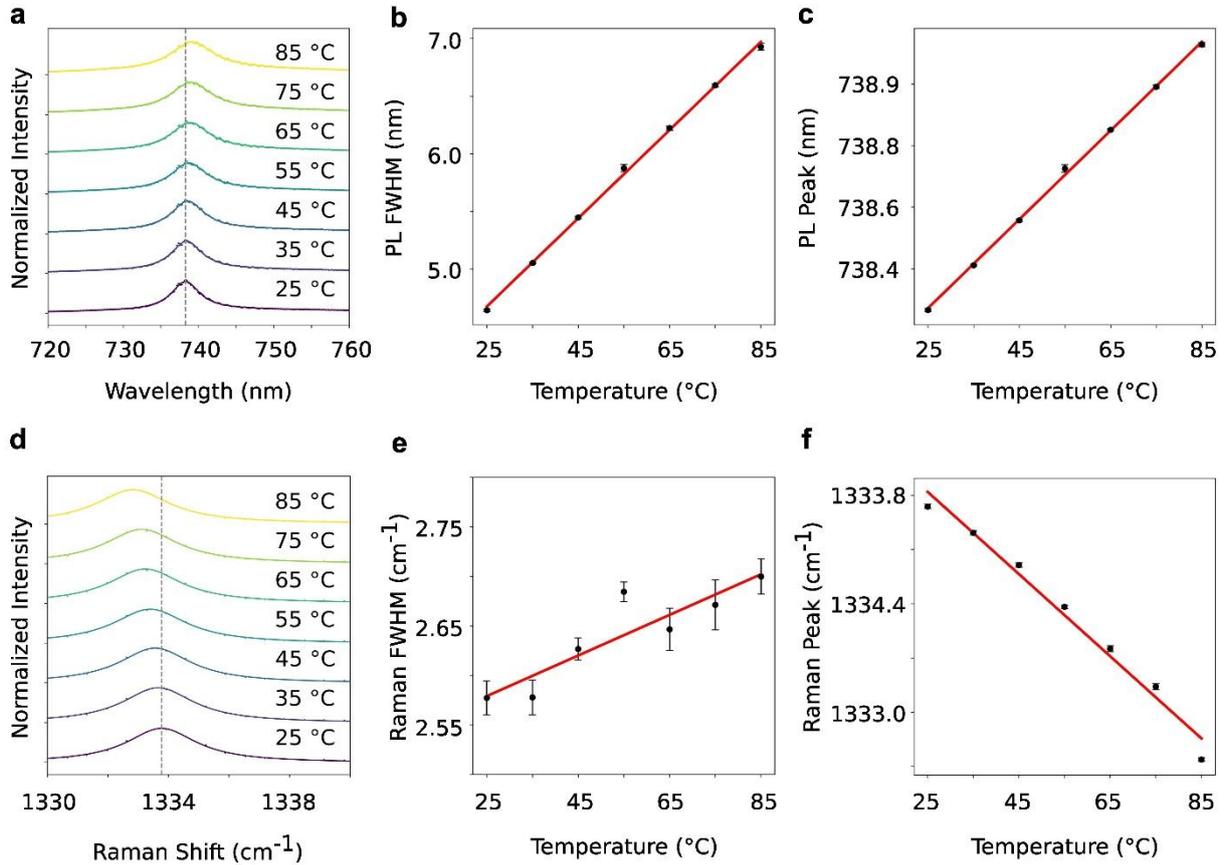

**Figure 2.** Temperature-dependent Photoluminescence (PL) and Raman spectra of diamonds hosting silicon-vacancy (SiV) centers. **a)** PL spectra (1-s acquisition time) of SiV centers at different temperatures. **b, c)** Temperature dependence of the full width at half maximum (b) and position (c) of the zero-phonon line (ZPL) of SiV centers. As temperature increases, the ZPL widens and shifts to longer wavelengths. **d)** Raman spectra (20-s acquisition time) of a representative diamond at different temperatures. **e, f)** Temperature dependence of the full width at half maximum (e) and position (f) of the diamond Raman peak. As temperature increases the Raman peak widens and shifts to lower wavenumbers. Each data point in (b), (c), (e), and (f) represents the mean value obtained from seven independent measurements performed on the same representative microdiamond. The error bars are the standard deviation of the measurements taken at each temperature. All spectral data were acquired using a 500-µW, 532-nm continuous wave (cw) laser.

**Figure 2** shows a condensed summary of the experimental PL and Raman data used to construct the MFR and 2SMFR models. Photoluminescence and Raman spectra acquired at different reference temperatures are displayed for a representative diamond (**Figure 2a** and **2d**), alongside the temperature-dependent behaviour of a selected subset of features: specifically, the SiV PL's and diamond Raman's linewidth and position (**Figure 2b, c** and **2e, f,** respectively). Note that the fit in **Figure 2e** shows relatively larger errors than those in **Figures 2b, c** and **f**. This further supports the idea of choosing multi-feature regression models over single-variable ones, as averaging and weighting across features can reduce the impact of noise and relative errors of individual features.

To build the MFR and 2SMFR models, the data is divided into two sets: a training dataset, and a test dataset containing data points not seen during training. The training set includes a subset of diamonds and associated temperature values, which are used to determine the predicted temperatures (cf. **§4.3, Methods**). A schematic of the training and testing workflow for the MFR algorithm is shown in **Figure 1d**; the workflow for the 2SMFR is conceptually analogous. Once trained, the model is applied to the test set. For each $i$-th entry, the MFR and 2SMFR algorithms generate predicted values $\tau_i$ for the temperature (as defined in Eqs. 1 and 3, respectively), which are then compared to the corresponding ground truth temperature values $T_i$ measured by the cryostat. Each model's accuracy and resolution are then assessed from the distribution of absolute errors $|\tau_i - T_i|$ across the test set (cf. **§4.4 Methods**).

To further optimize the performance of the MFR algorithm, we implemented an automated feature selection routine in which the model is trained and tested on every possible combination of features to identify the subset that yields the best performance (i.e., lowest prediction error and resolution). With 9 total features, this corresponds to $2^9-1=511$ non-empty feature subsets spanning sizes 1 through 9. This exhaustive approach is highly flexible, as different users may have instruments and experimental setups with varying technical specifications, allowing them to adapt the MFR model to rely more heavily on some features and less on others. For completeness, we remark that while computationally tractable in our case—only 9 features were considered here—exhaustive feature selection can become impractical, as the number of subsets increases exponentially with the number of features.

This explicit feature selection routine is not applied to the 2SMFR algorithm, which is based on Lasso regression. Lasso includes an L1 regularization penalty that can shrink some regression coefficients exactly to zero, thereby performing an implicit feature selection during model fitting. Lasso regression might therefore appear preferable in this context; however, its implicit feature selection can be unstable when predictors are highly correlated,[24] which is less of a concern for MFR.

In this work, we considered both MFR and 2SMFR, as each method shows distinct advantages compared to traditional techniques based on single-feature tracking (cf. **§2.2 below**). This allows users to choose the most suitable approach for their specific application requirements.

**2.2. Model characterization and benchmarking**

The primary goal of this work is to develop a practical and generalizable approach for micro/nanoscale temperature measurement that is directly applicable to real experimental settings. To this end, the MFR and 2SMFR algorithms are optimized, and their performance characterized and benchmarked using criteria specifically tailored to this objective—criteria that may differ from those used in similar studies.

Most optical micro/nanothermometers are secondary thermal sensors, meaning they must be calibrated against known temperature values—ideally in situ, under the same conditions in which they will ultimately be deployed. However, calibrating each sensor individually is often impractical and rarely feasible in real-world applications. To address this limitation, we adopt the train-test strategy introduced above: calibration is performed on the training set, while performance is benchmarked independently on the test set. The aim is to determine—once—the optimal calibration (Eqs. 1 and 3, **§4.3 Methods**) of the MFR and 2SMFR models, which, once trained and validated, can then be applied to any never-seen-before diamonds, not included in the original training or test datasets.

This framework leads us to consider two key aspects when optimizing and characterizing the models: *I)* generalized performance metrics, and *II)* efficiency of calibration.

*I)* The resolution and accuracy we report are generalized figures of merit (cf. **§4.4 Methods**), as they are computed on the independent test set rather than the training data used for model fitting. While this may seem trivial, it contrasts starkly with conventional approaches, where accuracy, resolution and sensitivity are often evaluated using the same dataset employed to train the model.[10, 23] This practice introduces what is known as optimistic bias, resulting in artificially inflated performance metrics. We note that this traditional approach is not inherently invalid, but it relies on the assumption that the same sensors used during calibration will also be used in the final application—a condition that is rarely met in practice. Accordingly, the generalized accuracy and resolution values we report here should be interpreted within this context and not directly compared to figures quoted in the literature using conventional methodologies.

*II)* Given the resource-intensive nature of the calibration process—which requires measurements across multiple diamonds and temperature set points—we systematically investigate how performance (i.e. accuracy and resolution) evolves with increasing amounts of calibration data. Our specific goal is to determine the smallest number of calibration diamonds and reference temperatures needed to achieve a desired level of precision, based on users' needs and application requirements.

The main results for the characterization and benchmarking of our models are presented in **Figure 3** and summarized in **Tables 1a** and **1b**. The two subpanels of Figure 3 display the accuracy and resolution of the multi-feature (MFR), two-step multi-feature (2SMFR) and the two best single-feature (SF) regression models as a function of the number of diamonds (**Figure 3a**) and of the number of temperature set points (**Figure 3b**) used for training. The reported accuracy and resolution values are the generalized values, in line with criteria *I* and *II* discussed above and with the definitions provided in the Methods section (**§4.4**). In these subpanels, data points indicate the mean accuracy across the available datasets, whereas the resolution—given by the standard deviation—is shown as shaded regions on the graph.

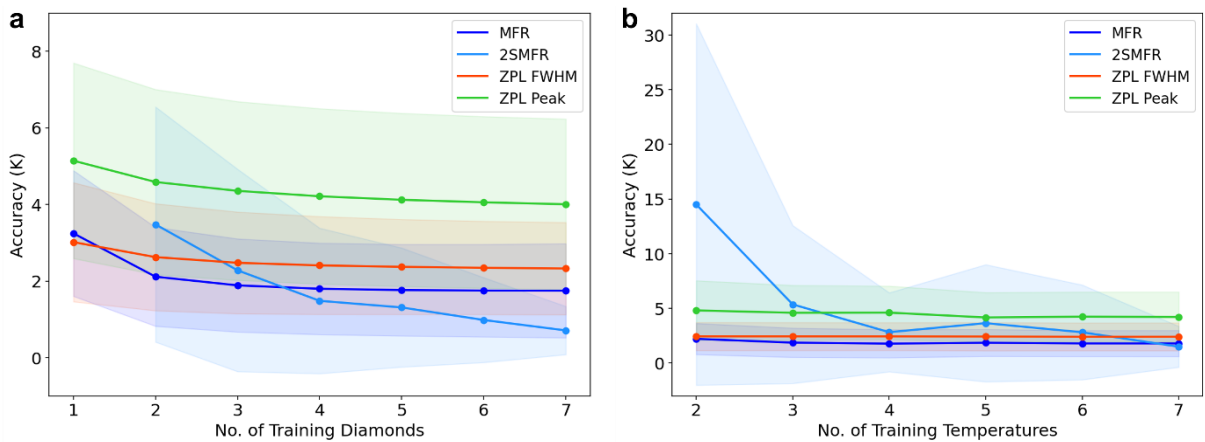

**Figure 3.** Performance of the tested nanothermometry single-feature (SF), multi-feature (MFR) and two-step multi-feature (2SMFR) regression models. The 2SMFR, best MFR model and the two best SF models are displayed. **a)** Accuracy and resolution as a function of the number of diamonds used for training. **b)** Accuracy and resolution as a function of the number of temperature set points used for training; for (b) calculations were done using 4 training diamonds. For both (a) and (b), values are displayed as data points on the graphs and resolution as shading around each corresponding accuracy value. Overall, the MFR model with features SiV's ZPL wavelength, SiV's linewidth and Raman linewidth (dark blue curve) and the 2SMFR model (light blue curve) outperform—they have lower

absolute errors—the best SF models (red and green curves). As expected, accuracy and resolution (of every model) improve as more diamonds and temperature set points are used to train the models.

Our dataset contains eight diamonds. To increase the sample size while maintaining statistical independence, we train and test the model on all possible independent combinations of diamonds for each chosen training set size. For example, if we wish to evaluate performance using three diamonds for training and the remaining five for testing, we calculate accuracy and resolution for every distinct 3-diamond training set and then average the results. In this example, the total number of such combinations would be given by the binomial coefficient $\binom{8}{3} = \frac{8!}{3!(8-3)!} = 56$.

**Table 1a: Accuracy and Resolution vs. No. of Training Diamonds**

|  | 1 | 2 | 3 | 4 | 5 | 6 | 7 |
|---|---|---|---|---|---|---|---|
| MFR | (3.2 ± 1.6) K | (2.1 ± 1.3) K | (1.9 ± 1.2) K | (1.8 ± 1.2) K | (1.8 ± 1.2) K | (1.8 ± 1.2) K | (1.7 ± 1.2) K |
| 2SMFR | – | (3.5 ± 3.1) K | (2.3 ± 2.6) K | (1.5 ± 1.9) K | (1.3 ± 1.6) K | (1.0 ± 1.1) K | (0.7 ± 0.6) K |
| ZPL FWHM | (3.0 ± 1.6) K | (2.6 ± 1.4) K | (2.5 ± 1.3) K | (2.4 ± 1.3) K | (2.4 ± 1.2) K | (2.3 ± 1.2) K | (2.3 ± 1.2) K |
| ZPL Peak | (5.1 ± 2.6) K | (4.6 ± 2.4) K | (4.4 ± 2.3) K | (4.2 ± 2.3) K | (4.1 ± 2.3) K | (4.1 ± 2.2) K | (4.0 ± 2.2) K |

**Table 1b: Accuracy and Resolution vs. No. of Training Temperatures**

|  | 2 | 3 | 4 | 5 | 6 | 7 |
|---|---|---|---|---|---|---|
| MFR | (2.2 ± 1.4) K | (1.9 ± 1.3) K | (1.8 ± 1.3) K | (1.9 ± 1.2) K | (1.8 ± 1.2) K | (1.8 ± 1.2) K |
| 2SMFR | (14.5 ± 16.5) K | (5.4 ± 7.2) K | (2.8 ± 3.6) K | (3.6 ± 5.4) K | (2.8 ± 4.3) K | (1.5 ± 1.9) K |
| ZPL FWHM | (2.4 ± 1.3) K | (2.4 ± 1.3) K | (2.4 ± 1.3) K | (2.4 ± 1.3) K | (2.4 ± 1.3) K | (2.4 ± 1.3) K |
| ZPL Peak | (4.8 ± 2.7) K | (4.6 ± 2.5) K | (4.6 ± 2.4) K | (4.2 ± 2.3) K | (4.2 ± 2.3) K | (4.2 ± 2.3) K |

**Table 1.** Accuracy and Resolution of the different models as a function of the number of training diamonds (1a) and of training temperature set points (1b). For Table 1b, the values are estimated assuming training with four diamonds.

Analysis of **Figure 3** and **Tables 1a**, **b** reveal several key findings.
We analyze first the results for the MFR algorithm. The MFR model consistently outperforms all SF models: regardless of the number of training diamonds or temperature set points, its accuracy, resolution, and associated dispersions are systematically lower (i.e., better) than those of the SF models. The best-performing MFR model achieves an accuracy of 1.7 K, representing a marked ~26% improvement over the 2.3 K obtained by the best-performing SF model. Both models exhibit comparable resolutions of approximately 1.2 K·Hz$^{-1/2}$. We note that these *absolute* values for accuracy and resolution are modest, as they represent generalized values (cf. point I above, and **§4.4** of the **Methods**). Nevertheless, the ~26% *relative* improvement in accuracy achieved by the MFR model over the best-performing SF model is significant— especially given that improved instrumentation could further reduce the absolute values.

Interestingly, the best performing MFR model only uses three features: SiV's ZPL wavelength, SiV's linewidth (FWHM) and Raman linewidth (FWHM). This is likely due to the fact that the measurement of these three physical quantities is less noisy than the others, or their noise averages out to produce lower absolute errors. Note that prior knowledge of which features would produce the best MFR model is not required, as our algorithm automatically employs the aforementioned explicit feature-selection routine that cycles through all possible permutations of feature combinations (511 in our case) to minimize the prediction error (cf. **§2.1**).

Second, we present the results for the 2SMFR algorithm. When trained with the maximum number of calibration diamonds (seven) and temperature set points (seven), the 2SMFR model achieves the best overall performance, with an accuracy of 0.7 K and a resolution of 0.6 K·Hz$^{-1/2}$. These *absolute* values are field-competitive given that they are generalized values. The 2SMFR accuracy of 0.7 K represents a relative improvement of ~59% (or a factor ~2.5×) compared to the 1.7 K accuracy of the MFR model, and ~70% (or a factor ≳3×) compared to the 2.3 K accuracy of the best SF model. Likewise, the 2SMFR resolution of 0.6 K·Hz$^{-1/2}$ corresponds to an improvement of ~50% (or a factor 2×) over both the MFR and the best SF models. While not being a strict requirement, we remark that a minimum of two training diamonds is required for the 2SMFR algorithm to work (cf. **§4.3.2**). Also, the 2SMFR model seems to require more training data than the MFR model to maximize performance and consistently starts to perform better when the training set includes at least four diamonds and four temperature set points.

This observation reflects a general trend: increasing the number of training diamonds and temperature set points enhances both accuracy and resolution across all models. For the 2SMFR model, increasing the number of training diamonds from two to seven reduces the accuracy error from 3.8 K to 1.5 K (an improvement of ~60%). For the MFR model, expanding from one to seven training diamonds lowers the accuracy error from 3.2 K to 1.7 K (~47% improvement). A smaller yet noticeable effect is seen for the best SF model, where the accuracy error decreases from 3.0 K to 2.3 K (~23% improvement). Resolution follows a similar pattern: for the MFR and SF models, resolution improves from 1.6 to 1.2 K·Hz$^{-1/2}$ (~25% improvement), while for the 2SMFR model it improves more substantially, from 4.4 to 2.1 K·Hz$^{-1/2}$ (~52% improvement).

We also observe similar trends for the number of temperature set points. To illustrate this, we present a representative case using four diamonds in the training dataset. Increasing the number of training temperature set points from two to seven reduces the accuracy error for the 2SMFR model from 14.5 K to 1.5 K (~90%) and for the MFR model from 2.2 K to 1.8 K (~22% improvement); conversely, there is no significant improvement for the best SF model for which the accuracy stays constant at a relatively large value, 2.4 K. Similarly, resolution improves going from 16.5 to 1.9 K·Hz$^{-1/2}$ (~88% improvement) for the 2SMFR model and from 1.4 to 1.2 K·Hz$^{-1/2}$ (~16% improvement) for the MFR model, while for the best SF model it remains unchanged at 1.3 K·Hz$^{-1/2}$.

The fact that increasing the number of training diamonds and/or temperature set points improves the performance of the models is expected—larger training datasets tend to average out outliers. However, the graphs reveal a particularly relevant trend for the goals of this work. In the case of the MFR model, using more than three training data points yields only marginal gains in accuracy and resolution. Specifically, increasing the training set from one to two diamonds improves accuracy from 3.1 K to 2.1 K—a substantial ~32% gain. Adding a third diamond reduces the error to 1.9 K, but this represents only a further ~10% improvement. Beyond three diamonds, the gains drop below 5%. A similar pattern emerges for the number of temperature set points: accuracy improves markedly when going from one to two, and from two to three calibration temperatures, but improvements taper off thereafter. This suggests that if one was to adopt the MFR algorithm, calibrating against more than just a few diamonds or temperature points may be unnecessary in time- or efficiency-critical applications.

The results for the 2SMFR algorithm reveal a more consistent improvement pattern. The largest gains in accuracy occur when increasing the number of training microdiamonds from two to three and from three to four, where accuracy improves from 3.5 K to 2.3 K and from 2.3 K to 1.5 K, respectively—each corresponding to roughly a 35% gain. Beyond this point, adding more training diamonds continues to improve accuracy, though more gradually, with each step yielding an additional ~13–25% improvement. A similar trend is observed with the number of temperature set points: the most dramatic gain occurs when increasing from one to two points, leading to a ~63% improvement. Further increases continue to enhance accuracy in a steady manner, with each additional point contributing a ~23–47% gain.

These observations and trends are validated from the analysis of **Figure 4**. The figure displays reference heatmaps for both the 2SMFR (**Figure 4a** and **4b**) and MFR models (**Figure 4c** and **4d**), showing how accuracy and resolution vary with any combination in the number of training microdiamonds and temperature set points. These visualizations show that performance improves systematically with increasing training set size, with the most pronounced gains occurring at the lower end of the range. The heatmaps also reveal how incremental additions of diamonds or temperature set points translate into diminishing, but still measurable improvements at higher values. In addition to confirming these performance trends, the heatmaps provide a practical reference for determining the minimum calibration requirements needed to achieve target accuracy and resolution in specific applications.

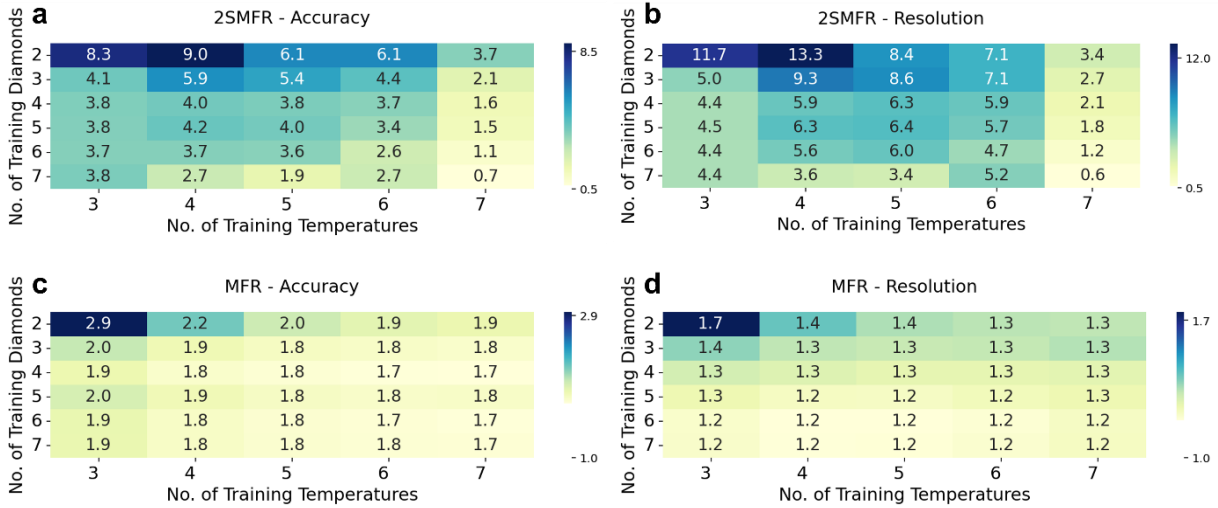

**Figure 4.** Accuracy and resolution heatmaps for the 2SMFR and MFR models. **a, b)** Accuracy (a) and resolution (b) values of the 2SMFR model for different combinations of number of training diamonds and temperature set points. **c, d)** Accuracy (a) and resolution (b) values of the best MFR model for different combinations of number of training microdiamonds and temperature set points.

The final metric we consider is the relative sensitivity, an important indicator of performance for any thermometry technique. Relative sensitivity is an absolute measure that specifies the smallest detectable change in temperature for a given sensor. Following the criteria described in §4.4, we calculate the relative sensitivity of each method using the case with four training diamonds and seven temperature set points as a reference. The relative sensitivities are as follows: $4·10^{-3}$ K$^{-1}$ fort he MFR model, $4·10^{-2}$ K$^{-1}$ for the 2SMFR model, $6·10^{-3}$ K$^{-1}$ for the SF model based on the FWHM of the SiV ZPL, and $1.7·10^{-5}$ K$^{-1}$ for the SF model based on the SiV's ZPL wavelenght. Therefore alongside exhibiting the best accuracy and resolution the 2SMFR model also display the highest sensitivity (roughly by a 7× factor) compared to the best SF model. The sensitivity oft he MFR model is comparable to that of the best SF model.

Before concluding, we note two practical considerations. First, our experiments and models assumed a linear relationship between each feature and temperature. This assumption is reasonable within the limited range studied here (25–85 °C), but may break down over broader ranges, with the effect of reducing the predictor's accuracy. Favorably, in such cases accuracy can be recovered by pre-linearizing the data using their known functional dependence on

temperature—a relationship typically known for the spectroscopy features of diamond, its color centers, and in general of other optical thermometers. Second, regardless of the model used—traditional SF or the proposed MFR/2SMFR models—predictions beyond the training set's temperature range should be treated with caution, as extrapolation outside of it can lead to unreliable results.

## 3. Conclusion

In this work, we demonstrated an all-optical thermometry technique based on fluorescent diamonds containing SiV centers, where the thermo-sensitive observables are the SiV photoluminescence and the Raman signal of the diamond host. To analyze the experimental data, we employed two algorithms: a multi-feature regression (MFR) model and a two-stage multi-feature regression (2SMFR) model. Both were trained on calibration data from a subset of diamonds and then used to predict temperatures for previously unseen ones.

Our goal was deliberately practical: to establish that these algorithms can automatically identify the optimal combination of observables (features) and determine the minimal number of calibration temperatures required to achieve target accuracy and resolution. With only 3–4 calibration diamonds and temperature set points, the MFR model predicts temperatures with an accuracy of 1.7 K and a resolution of 1.2 K·Hz$^{-1/2}$. This corresponds to roughly a 26% improvement in accuracy over traditional single-feature models, while maintaining a comparable resolution.

The 2SMFR algorithm further improves performance, though at the cost of larger calibration sets. When trained with seven microdiamonds and seven temperature set points, it achieves an accuracy of 0.7 K (representing ~59% improvement relative to MFR and ~70% relative to the best single-feature model) and a resolution of 0.6 K·Hz$^{-1/2}$ (~50% better than both the MFR and best single-feature models). The 2SMFR model also achieve a sensitivity of 0.04 K$^{-1}$ which correspond to a sevenfold improvement over that of the best SF model (0.006 K$^{-1}$).

Thus, compared to traditional thermometry techniques that monitor a single variable, our approach based on multi-feature regression can achieve accuracies, resolutions and sensitivities that are notably better by factors ≳3×, 2× and 7×, respectively. Taken together, these results demonstrate that multi-feature approaches offer a powerful advance over traditional single-feature methods, enabling higher accuracy, resolution and sensitivity with flexible calibration requirements. The techniques introduced here are therefore a powerful addition to the variety of existing all-optical thermometry techniques and provide a broadly applicable framework for any optical sensor with at least two temperature-dependent observables.

## 4. Materials and Methods

### 4.1. Sample preparation

SiV⁻ diamonds were purchased from commercially available sources (Adámas Nanotechnologies). The diamonds (approximately 1 μm in size) were dispersed in IPA and sonicated for 2 minutes to ensure uniform dispersion. The solution was then drop cast on a pre-cleaned silica substrate (0.5×0.5 cm²) for characterization and allowed to dry on a hotplate at 200 °C for 20 minutes to evaporate the solvent and enhance the adhesion of diamond to the substrate. During high-pressure, high-temperature synthesis, disordered *sp²* carbon can form on the diamond surfaces and then coalesce into continuous graphitic layers. To remove these graphitic residues and regenerate a pristine *sp³*-terminated surface, the microdiamond sample was further oxidatively annealed in ambient air at 550 °C for 2 h in a tube furnace (Lindberg Blue Mini-Tube Furnace), selectively oxidising the *sp²* carbon while leaving the underlying *sp³* diamond intact.[25] **Figure S1** (cf. **Supporting Information)** shows the Raman spectrum of the diamond after tube-furnace treatment. Eight diamonds exhibiting uniform morphology were selected for the experiment.

### 4.2. Experimental Setup

Raman and photoluminescence (PL) spectra were acquired using a lab-built confocal microscopy setup (cf. **Supporting Information, Figure S2**). The sample was mounted on a high-precision temperature controller (Microoptik MHCS600) using silver conductive paste to improve thermal conductivity between the temperature controller and the sample. A continuous wave (cw) 532-nm laser with excitation power of 500 μW was focused on the SiV diamond on the sample through a high numerical-aperture objective (NA=0.7; MY100X-806, 100×; Thorlabs). The emission signal from the microdiamond was back-collected using the same objective and directed to a spectrometer (ANDOR-SR-500i) via a 30R/70T cube beam splitter and a long pass filter to isolate the PL and Raman signal from the excitation laser (Semrock, LP 561). Photoluminescence (PL) and Raman spectra were recorded for each of the eight diamonds at specific temperatures (25–85 °C) at 10 °C increments, with a 10-minute stabilization period at each step to ensure thermal equilibrium was reached before acquiring the spectral data. This procedure was repeated at each temperature point to maintain consistency and reliability in the measurements. The experimental value of each set temperature as measured by the controller was considered the 'true' temperature value for the purpose of determining accuracy and resolution of our method.

### 4.3. Regression models

Most machine learning (ML) approaches aim to predict outcomes based on past observations, using a training dataset to construct a predictive model and a test dataset to evaluate its accuracy on known data before applying it to unknown datasets.

*4.3.1. Multi-feature regression (MFR)*

The first model we use in this work, is a multi-feature linear regression model that predicts the temperature $T$ from the simultaneous measurement of $n$ features, $\mathbf{x}_i = [x_{i1}, x_{i2}, \ldots, x_{in}]$ such as photoluminescence intensity, zero-phonon-line (ZPL) wavelength, and full width at half maximum (FWHM) of color centers in nanodiamonds and the Raman signal of the diamond host.

The predicted temperature, $\tau_i$, determined by the model is thus:

$$\tau_i = \sum_{j=0}^{n} w_{ij} x_{ij} = w_{i0} + w_{i1} x_{i1} + w_{i2} x_{i2} + \cdots + w_{in} x_{in} \tag{1}$$

where the extended input vector $\mathbf{x}_i = [1, x_{i1}, x_{i2}, \ldots, x_{in}]$ includes a constant term to account for the intercept $w_{i0}$, and $i = 1, 2, \ldots, N$ indexes the dataset ($N$ is the total number of datasets).

The weights $\mathbf{w}_i = [w_{i0}, w_{i1}, \ldots, w_{in}]$ are determined from the training data by minimizing the least squares error (LSE, or L2-norm) cost function:

$$LSE = \sum_{i=1}^{N} (\mathbf{w}_i \cdot \mathbf{x}_i - \tau_i)^2 \tag{2}$$

Note that we analyze the data using Python's open-source machine learning library scikit-learn.[26] All input features are normalized via MinMax scaling to the [0, 1] range, ensuring that variables with larger numerical ranges do not overshadow smaller ones. This normalization allows the regression analysis to capture their true relative influence.

*4.3.2. Two-stage multi-feature regression (2SMFR)*

The second model used in this work is a two-stage regression algorithm that leverages both photoluminescence (PL) and Raman spectra to predict temperature.

For each sample, the PL spectrum is z-scored across wavelengths, i.e., for each spectrum we subtract its mean intensity and divide by its standard deviation. This normalization removes baseline offsets and variance scaling, eliminating absolute intensity differences between spectra—for example, those arising from laser power fluctuations, detector gain changes, or alignment drift—and forces the model to focus on relative spectral shape. Note that this means absolute PL intensity is no longer a usable feature in the model.

From the z-scored PL spectra, we extract $k$ latent scores using partial least squares (PLS) regression (scikit-learn[26]), with the value for $k$ chosen to maximize covariance with temperature. The Raman data from the same acquisition contributes three scalar features: peak position, peak height normalized by the PL maximum and unnormalized full width at half maximum.

The PLS-derived latent scores and the Raman scalars are concatenated to form the feature matrix. Each column is standardized (zero mean, unit variance) within the dataset using StandardScaler. A Lasso regression model is then trained on the standardized features to produce a base temperature estimate $\tau_{0,i}$.

To correct systematic biases in $\tau_{0,i}$, we fit a univariate residual calibrator $r(\tau_{0,i})$. This is implemented through the following steps. Expanding $\tau_{0,i}$ into a B-spline basis (SplineTransformer); standardizing without centering; fitting a ridge regression model (RidgeCV) to predict the residuals $\tau_{0,i} - T_i$ (where $T_i$ is the true temperature as measured by the cryostat).

The final prediction is then:

$$\tau_i = \tau_{0,i} - s \cdot r(\tau_{0,i}) \tag{3}$$

where $s$ is a learned scalar coefficient. Because our dataset is relatively small, the calibration step is performed on the training dataset rather than on an independent validation set. To mitigate the potential bias from this choice, we use cross-validated calibration within the training set: the training data is internally split into folds, with some folds used to fit the base model and the remaining folds used solely for calibrating it. All possible training/calibration fold assignments are used, ensuring statistical independence between base-model fitting and calibration in each split. Since at least one fold is needed for training and one for calibration, this method requires a minimum of two diamonds in the training set.

### 4.4. Data analysis

To characterize the performance of our approach we measure accuracy, resolution and sensitivity with the following considerations. The values we determine for resolution, accuracy, and sensitivity are what we refer to as *generalized*. This approach involves first training each model—the proposed MFR and 2SMFR models, as well as the SF model used as benchmark—on known calibration (training) datasets. We then test the models on separate, previously unseen (test) nanothermometry data to evaluate how well they predict the true temperature. Unlike traditional approaches, these generalized figures of merit are evaluated on data not used during training, and therefore typically yield larger (i.e., worse) values (see main text).[10, 23]

For the purpose of our estimates, the temperature values measured by the cryostat are considered the true reference. Strictly speaking, a rigorous—albeit impractical—definition would require exact knowledge of the absolute temperature, rather than relying on measurements from a reference instrument.

***Accuracy:*** Accuracy is defined as the difference between the measured (average) value and the 'true' value of an observable—in this case, temperature. In our characterization, the measured value refers to the temperature predicted by the model (MFR, 2SMFR or SF), obtained by fitting the experimental data. This predicted value is then directly compared to the reference temperature provided by the cryostat.

***Resolution***: In micro-/nanothermometry, resolution is typically defined as $\sigma\sqrt{t_m}$ where $\sigma$ is the standard deviation of the measured feature (e.g., intensity, ZPL, FWHM) and $t_m$ is the integration time. In our case, we compute $\sigma$ as the standard deviation of the absolute differences between the true temperatures and the model-predicted temperatures (from either the MFR/2SMFR or control SF models). The resolution is then obtained by multiplying this standard deviation by $\sqrt{t_m}$. Alternatively, resolution can be evaluated by using either the largest (worst) or smallest (best) value of $\sigma$ across all tested temperatures.

***Sensitivity:*** The relative sensitivity of a nanothermometry technique is defined as $|(\partial O/\partial T)/O|$, where $O$ is the measured observable and $T$ is the temperature. For each control single-feature (SF) model, sensitivity values are obtained directly using this definition.

In contrast, the sensitivity of the MFR and 2SMFR models cannot be calculated directly in the same manner. For the MFR model, however, the general definition can still be applied, provided it is adapted to account for multiple features. Specifically, the relative sensitivity can be estimated as a weighted linear combination of the individual sensitivities associated with each SF model (calculated as per the direct definition above). Specifically, if the MFR model uses $n$ features, the sensitivity of the $i$-th dataset is:

$$S_{MFR,i} = \sum_{j=1}^{n} \alpha_{ij} \left(\frac{\partial x_{ij}}{\partial T}\right)/x_{ij} \qquad (4)$$

where the individual sensitivities for each feature, $s_{ij} = (\partial x_{ij}/\partial T)/x_{ij}$, are weighted by the corresponding coefficient $\alpha_{ij}$. These coefficients $\alpha_{ij}$ are obtained from the coefficients $w_{ij}$ in Equations 1 or 3, via normalization: $\alpha_{ij} = w_{ij}/\sum_{j=1}^{n} w_{ij}$. The index $j$ runs over the number of features used in the MFR model, while the index *i* refers to the dataset of the corresponding *i*-th diamond. For the 2SMFR model, this approach is difficult to apply because the PLS transformation—which combines correlated features into orthogonal components to best explain variation in the response—makes it impractical to extract the relative weights needed for the sensitivity of each feature. We therefore adopt an empirical definition: we evaluate the model's output, $O$, at two different temperatures, $T$ and $T + \Delta T$, and calculate the relative sensitivity of the *i*-th dataset simply as:

$$S_{2SMFR,i} = \frac{O_i(T + \Delta T) - O_i(T)}{O_i(T)\Delta T} \tag{5}$$

In both Equations (4) and (5), the reported sensitivity corresponds to the average value across all datasets *i*.

# Supporting Information

**Machine Learning Based Optical Thermometry Using Photoluminescence and Raman Spectra of Diamonds Containing SiV Centers**

*Md Shakhawath Hossain,[†] Dylan G. Stone,[##] G. Landry,[##] Xiaoxue Xu,[††] Carlo Bradac[##,*] and Toan Trong Tran [†, *]*

[†] School of Electrical and Data Engineering, University of Technology Sydney, Ultimo, NSW, 2007, Australia.

[##] Department of Physics & Astronomy, Trent University, 1600 West Bank Dr., Peterborough, Ontario K9L 0G2, Canada

[††] School of Biomedical Engineering, University of Technology Sydney, Ultimo, NSW, 2007, Australia.

*Corresponding author: carlobradac@trentu.ca

*Corresponding author: trongtoan.tran@uts.edu.au

The Supporting Information information includes:

**Supporting Information Figure S1:** Raman spectra of a SiV⁻ diamond after annealing in tube furnace fitted with single lorentzian function

**Supporting Information Figure S2:** Schematic of the experimental setup for Raman and photoluminescence (PL) measurements.

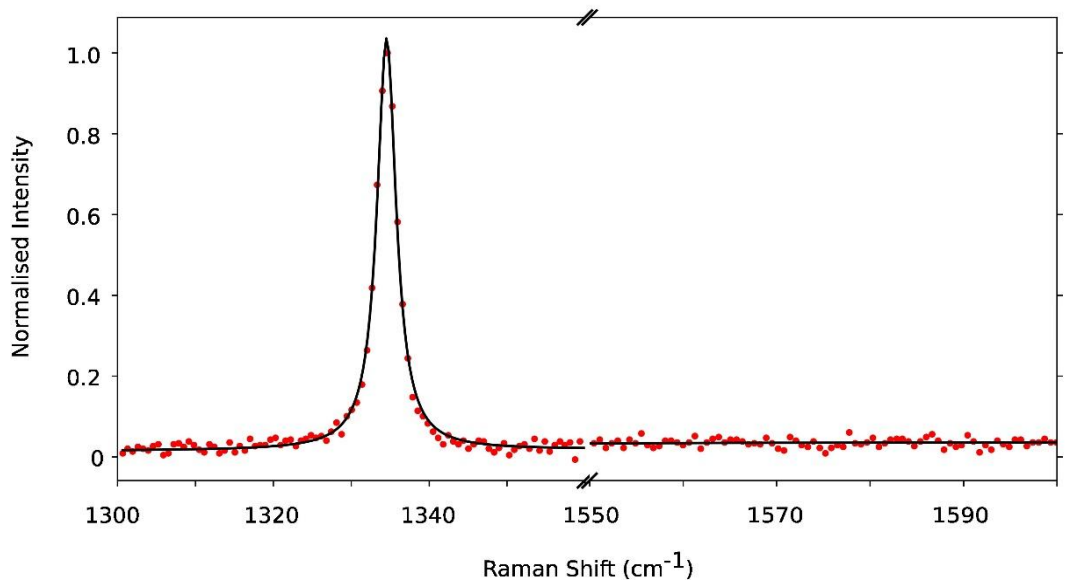

**Figure S1** Raman spectra of a SiV¯ diamond after annealing in tube furnace fitted with single lorentzian function. Raman peak is at ~1335 cm$^{-1}$. The spectrum is taken with 20s acquisition time and 500 µW excitation power through the objective. The absence of a G-band feature around 1580 cm$^{-1}$ confirms that no graphitic layer is present on the microdiamond.

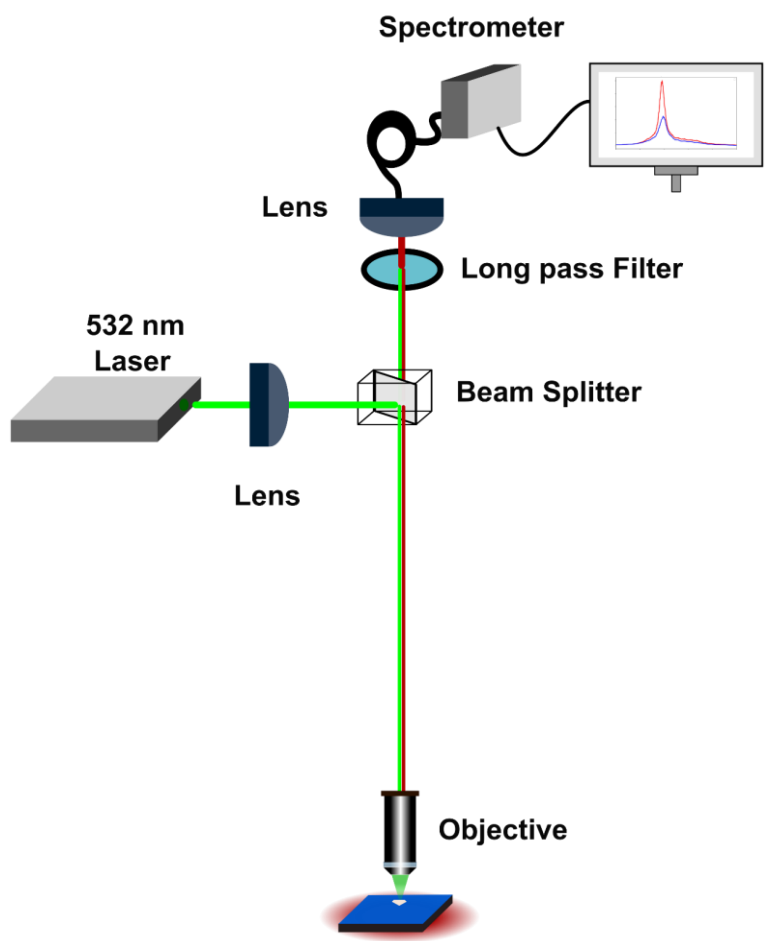

**Figure S2** Schematic of the experimental setup for Raman and photoluminescence (PL) measurements. A 532-nm continuous-wave (cw) laser is focused onto the sample via a high numerical aperture (NA) objective. The emitted signal is directed through a 30R/70T beam splitter and a 561-nm long-pass filter before being collected by a spectrometer equipped with a charge-coupled device (CCD) camera.